\documentclass[aps,prd,twocolumn,reprint,preprintnumbers,floatfix,nofootinbib,superscriptaddress]{revtex4-1}

\input{header}
\usepackage{ulem}

\def\carleton{Department of Physics, Carleton University, Ottawa, ON K1S 5B6, Canada }

\newcommand{\mgr}{m_{3/2}}
\newcommand{{\bivo}}{{bi$\nu$o}}

\begin{document}
\title{Discovering the Origin of Neutrino Masses at SHiP}

\author{Seyda Ipek}
\email{sipek@physics.carleton.ca}
\author{Douglas Tuckler}
\email{dtuckler@physics.carleton.ca}
\affiliation{\carleton}

\begin{abstract}
In $U(1)_R$ extensions of supersymmetric models, the bino and its Dirac partner, the singlino, can play the role of right-handed neutrinos. The bino and the singlino form a pseudo-dirac pair, dubbed the `bi$\nu$o', which can generate Standard Model neutrino masses via the inverse seesaw mechanism. We investigate the prospects for detecting long-lived bi$\nu$os at SHiP, where GeV scale bi$\nu$os can be copiously produced in the decays of mesons. We show that SHiP can probe new regions of parameter space that are complementary to searches for the lepton flavor-violating decay $\mu \to e \gamma$. This scenario provides a well-motivated benchmark for future experiments of a right-handed neutrino that mixes with all Standard Model neutrinos, and is directly related to the generation of neutrino masses.
\end{abstract}

\maketitle

\section{Introduction}

The observation of neutrino oscillations indicates that at least two of the Standard Model (SM) neutrinos have non-zero masses. Measurements of atmospheric, reactor, solar, and accelerator neutrinos have determined the neutrino mass differences and mixing angles to be \cite{Gonzalez-Garcia:2021dve}
\begin{align}\label{eq:data}
&\Delta m_{21}^2 \simeq 7.4 \times 10^{-5} \ \text{eV}^2, \  \ |\Delta m_{31}^2| \simeq 2.5 \times 10^{-3} \ \text{eV}^2,   \nonumber \\
&\sin^2\theta_{12} \simeq 0.3, \ \sin^2\theta_{23} \simeq 0.45, \ \sin^2\theta_{13} \simeq 0.022 \ .
\end{align}
In the SM, neutrinos are massless and an explanation of non-zero neutrino masses requires beyond-the-SM (BSM) physics. A simple way to generate neutrino masses is to introduce Majorana fermions that are SM gauge singlets called right-handed neutrinos which lead to a suppression of SM neutrino masses via the seesaw mechanism \cite{Minkowski:1977sc,Gell-Mann:1979vob,Mohapatra:1979ia,Yanagida:1979as}. In this mechanism the light neutrino masses are inversely proportional to the Majorana mass and hence, a very heavy scale, $M\sim 10^{16}$~GeV, is needed to explain the smallness of SM neutrino masses.  In comparison, in the Inverse Seesaw (ISS) mechanism right-handed neutrinos are pseudo-Dirac fermions, with both Majorana and Dirac masses \cite{Mohapatra:1986aw,Mohapatra:1986bd,Branco:1988ex}. In this case, light neutrino masses are \emph{proportional} to the (small) Majorana mass.

It has been shown that the ISS mechanism can be realized in a $U(1)_R$-symmetric minimal supersymmetric SM (MSSM)~\cite{Coloma:2016vod}. In $U(1)_R$-symmetric MSSM gauginos are necessarily pseudo-Dirac fermions. The Dirac gaugino masses are produced via super-soft terms while the Majorana masses, proportional to the gravitino mass $\mgr$, can be produced via anomaly mediation. The pseudo-Dirac bino can be considered as a pseudo-Dirac right-handed neutrino, generating the neutrino masses. (The model is described in more detail in the next section.) Such a bino is called ``{\bivo}" to stress the neutrino connection. Light neutrino masses in this model are proportional to the ratio of the gravitino mass over the messenger scale $\Lambda_M$. Low energy observables like BR$(\mu\to e \gamma)$ give a constraint of $\Lambda_M\gtrsim 35$~TeV. For $\Lambda_M\sim O(100~{\rm TeV})$, generating the correct neutrino mixing parameters requires $\mgr \sim O(10~{\rm keV})$.

The {\bivo} model also provides rich collider phenomenology~\cite{Gehrlein:2019gtk, Gehrlein:2021hsk}. Since the {\bivo} mixes with neutrinos, it decays to a mixture of leptons, quarks and missing energy. If the {\bivo} is heavier than the weak scale, $M_{\tilde{B}}\gtrsim 90~$GeV, and the messenger scale is not too high, $\Lambda_M<10^8$~TeV, the {\bivo} decays promptly. However, if the {\bivo} is light, it needs to decay via off-shell $W/Z$ or Higgs to a three-body final state, making it a long-lived particle even for $\Lambda_M=100~$TeV. In~\cite{Gehrlein:2021hsk} a long-lived {\bivo} signal at MATHUSLA, FASER and CODEX-b was studied for $M_{\tilde{B}}>1$~GeV.

These earlier collider studies relied on a certain choice of the supersymmetric mass spectrum: the {\bivo} is the next-to-lightest supersymmetric particle (NLSP) and all other gauginos are heavier than the squarks. In this scenario, the {\bivo} is produced via squark decays. Hence, the constraints are cast on a combination of the squark mass, the {\bivo} mass and the messenger scale. However, if the squark masses are beyond the reach of high-energy colliders like the LHC, producing the {\bivo} in large quantities could be challenging.

Because the {\bivo} mixes with SM neutrinos, it can be produced in any process where a SM neutrino is produced. In particular, if they are light enough, {\bivo}s can be copiously produced in the decays of mesons at high energy beam dump experiments.\footnote{Neutralino decays in R-parity violating (RPV) MSSM can also be probed at experiments like SHiP, see~\cite{deVries:2015mfw}. The model we describe here is distinctly different than generic RPV.} In addition, {\bivo}s can become long-lived even for relatively low messenger scales. In this paper, we investigate the prospects for GeV-scale {\bivo}s at beam dump experiments such as SHiP. We show that the large production rate of {\bivo}s from meson decays enables the probe of new regions of parameter space that are not excluded by $\mu \to e \gamma$ and Big Bang Nucleosynthesis (BBN) constraints. Currently, long-lived particle experiments like SHiP can set the leading exclusion limits for {\bivo} masses in the $\sim1-5$ GeV range. Our results are summarized in Fig.~\ref{fig:sensitivity}.

This paper is organized as follows. In Sec.~\ref{sec:model} we introduce the model and fix our notation.  In Sec.~\ref{sec:Ship} we briefly describe the SHiP experiment and discuss meson production in the proton beam dump. The production and decay phenomenology of the {\bivo} is described in Sec~\ref{sec:pheno}. The sensitivity of SHiP is discussed in Sec.~\ref{sec:sensitiviy}. We conclude in Sec.~\ref{sec:conclusion}.

\section{Model}\label{sec:model}
In this section we summarize the relevant parts of the model we study. The details can be found in \cite{Coloma:2016vod, Gehrlein:2019gtk}. 

In $U(1)_R$-symmetric MSSM, a global $U(1)_R$ is imposed on the supersymmetric sector. The SM particles are not charged under this symmetry while the supersymmetric partners carry $+1$ $U(1)_R$ charges. We work with a modified version of this model where the global symmetry is instead $U(1)_{R-L}$, with $L$ being the lepton number. Due to this global symmetry, gauginos cannot be Majorana fermions. In order to give Dirac masses to gauginos, adjoint partners with $U(1)_{R-L}$ charges of $-1$ are introduced. For example, for the bino, $\tilde{B}$, there will be a superfield $\Phi_S$ whose fermionic component $S$, \emph{singlino}, is a SM singlet with $-1$ $U(1)_{R-L}$ charge. (Similarly, there is a tripletino and an octino as Dirac partners to the weakinos and the gluino respectively.) We only focus on the bino, lepton and Higgs superfields here as they are the relevant particle content for generating the neutrino masses.

We assume supersymmetry is broken in a hidden sector, via both $F-$ and $D-$terms and is mediated to the visible sector at a messenger scale $\Lambda_M$. Furthermore, as any global symmetry, $U(1)_{R-L}$ must be broken due to gravity. At the end, bino will have both a Dirac~\cite{Fox:2002bu} and a Majorana mass~\cite{Randall:1998uk, Giudice:1998xp}: 
\begin{align}
    M_{\tilde{B}} = c_i \frac{D}{\Lambda_M}\,,\quad m_{\tilde{B}} = \frac{\beta(g_Y)}{g_Y}m_{3/2}\,,
\end{align}
where $c_i$ is an $O(1)$ coefficient and $\mgr = \sum (F_i^2 + D_i^2/2)/\sqrt{3}M_{\rm Pl}^2$ is the gravitino mass.

In~\cite{Coloma:2016vod} it was shown that the following dimension-5 and dimension-6 operators
\begin{equation}\label{eq:neutrinomasses}
\frac{f_i}{\Lambda_M^2} \int d^2\theta W^\prime_\alpha W^\alpha_{\tilde{B}} H_u L_i \ , \ \frac{d_i}{\Lambda_M} \int d^4\theta \phi^\dagger \Phi_S H_u L_i\,,
\end{equation}
can generate two non-zero neutrino masses via the inverse seesaw mechanism. (Here $\phi = 1+\theta^2 \mgr$ is the conformal compensator.) Together with the mass terms for the bino, these interactions lead to the following Lagrangian
\begin{equation}\label{eq:LagMix}
\mathcal{L} \supset  M_{\tilde{B}} \tilde{B}S + M_{\tilde{B}}\tilde{B}\tilde{B} + f_i \frac{M_{\tilde{B}},}{\Lambda_M} \ell_i h_u \tilde{B} + d_i \frac{m_{3/2}}{\Lambda_M} \ell_i h_u S\,,
\end{equation}
where $f_i$ and $d_i$ are determined by the neutrino mass differences as
\begin{align}\label{eq:mixing}
f_i \simeq \begin{pmatrix} 0.35\\ 0.85\\ 0.35 \end{pmatrix}\,, \quad
d_i \simeq \begin{pmatrix} -0.06 \\ 0.44\\0.89  \end{pmatrix}.
\end{align}
After electroweak symmetry breaking (EWSB) the light neutrino masses are given by 
\begin{equation}\label{eq:masses}
m_1 = 0 , \ m_2 = \frac{m_{3/2}v^2}{\Lambda_M^2}(1-\rho), \ m_3 = \frac{m_{3/2}v^2}{\Lambda_M^2}(1+\rho),
\end{equation}
where $\rho \simeq 0.7$ is determined by the neutrino mass splittings in Eq.~(\ref{eq:data}). Note, that the coupling of $S$ to SM particles induced after electroweak symmetry breaking are proportional to the small gravitino mass $\mgr \sim\ \mathcal{O}$(1-10 keV), and will not play a phenomenological role. Therefore, we only focus on $\tilde{B}$ which we will call the ``\bivo'' for the rest of this paper.

\section{The SHiP Experiment}\label{sec:Ship}

The Search for Hidden Particles (SHiP) experiment is a proposed proton beam dump experiment that uses the 400 GeV proton beam at the CERN Super Proton Synchrotron (SPS) accelerator to search for long-lived particles \cite{Alekhin:2015byh,SHiP:2015vad,SHiP:2021nfo}. The proton beam extracted from the SPS will be dumped onto a high density target and provides 2$\times10^{20}$ protons-on-target in 5 years of operation. The entire SHiP experiment consists of a high density target followed by a hadron absorber, a muon shield, and a neutrino detector which all together have a length of $\ell_\text{sh} = $ 64 m. Immediately following the neutrino detector is a  decay volume with a length $\ell_\text{decay} = 50$ m, and a Hidden Sector Decay Spectrometer (HSDS) to detect the decay products of long-lived hidden sector particles. 

Long-lived particles with MeV-GeV masses can be produced from the decays of mesons. Because of the high energy of the proton beam and high density of the target, a large number of mesons will be produced in the beam dump when the incoming proton beam collides with the target. The production of kaons and heavy flavor mesons at SHiP has been previously studied in \cite{Alekhin:2015byh,SHiP:2018xqw,Gorbunov:2020rjx}. 

To determine the number of charged kaons produced at SHiP, we use the kaon production fractions from \cite{Gorbunov:2020rjx}. With a \texttt{GEANT4} simulation that takes into account production and propagation of kaons, it was found that  $\sim8$ $K^+$s and $\sim 3.5$ $K^-$s are produced per proton-on-target. About half of these are absorbed in the target, and a large fraction of the remaining kaons decay at rest. The resulting final states are isotropic and only a small number of the decay products will have trajectories that are in the direction of the HSDS.  Therefore, kaons that are stopped and decay at rest are neglected. The usable kaons are those that decay in-flight and it was found that 0.29 $K^+$ and 0.07 $K^-$ are produced per proton-on-target which can be used for the production of BSM particles.

The number of heavy flavor mesons produced at SHiP is given by \cite{Alekhin:2015byh,SHiP:2018xqw}
\begin{equation}\label{eq:Nmeson}
N_M = N_\text{POT}\times (2 \times X_{\bar{q}q} \times f^q_\text{cascade}) \times f(q\to M)\,,
\end{equation}
where $N_\text{POT} = 2\times 10^{20}$ is the number of protons-on-target, $X_{\bar{q}q}$ is the quark-antiquark production fraction, i.e. the probability of producing a $q\bar{q}$ pair in $pp$ collisions, $f^q_\text{cascade}~(q = c,b)$ is a cascade enhancement factor that takes into account secondary meson production in the target, and $f(q \to M)$ is the meson production fraction -- the probability that a quark $q$ will hadronize into a meson $M$. Values for these parameters are given in Tab. \ref{tab:params} and detailed discussions can be found in \cite{CERN-SHiP-NOTE-2015-009,SHiP:2018xqw}. Note that we only show values for $D^\pm_{(s)}$ and $B^\pm_{(c)}$ since these have the largest branching ratios to {\bivo}s above the kaon mass.

The total number of charged mesons produced at SHiP with $N_\text{POT} = 2\times 10^{20}$ are
\begin{align}
\begin{split}
N_{K^+} =&5.8\times 10^{19} ,~~~N_{K^-} = 1.4\times 10^{19}\,, \\
N_{D^\pm} =&3.2\times10^{17} ,~~~N_{D_s^\pm} =1.4\times10^{17}\,,\\
N_{B^\pm} =&4.5\times10^{13},~~~ N_{B_c^\pm} =2.8\times10^{11} \,,
\end{split}
\end{align}
where we have saturated the upper bound of $f(b \to B_c^\pm)$ to calculate $N_{B_c^\pm}$. Because of the large number of mesons produced, SHiP can have remarkable sensitivity to MeV-GeV-scale BSM particles.

\begin{table}
\begin{ruledtabular}
\begin{tabular}{ l l l l   }
$f(c \to D^\pm)$ & 0.207 & $X_{\bar{c}c}$ & $1.7 \times 10^{-3}$\\
$f(c \to D_s^\pm)$ & 0.088 & $X_{\bar{b}b}$ & $1.6 \times 10^{-7}$\\
$f(b \to B^\pm)$ & 0.417 & $f^c_\text{cascade}$& 2.3 \\
$f(b \to B_c^\pm)$ &  $\leq 2.6 \times 10^{-3}$& $f^b_\text{cascade}$ & 1.7
\end{tabular}
\end{ruledtabular}
\caption{\label{tab:params} Meson production fractions \cite{Graverini:2133817}, $q\bar{q}$ production fractions \cite{HERA-B:2007rfd,Lourenco:2006vw}, and cascade enhancement factors \cite{CERN-SHiP-NOTE-2015-009} for the heavy flavor mesons that are most relevant for {\bivo} production.}
\end{table}


\begin{figure*}[t]
    \centering
        \includegraphics[width=0.492\textwidth]{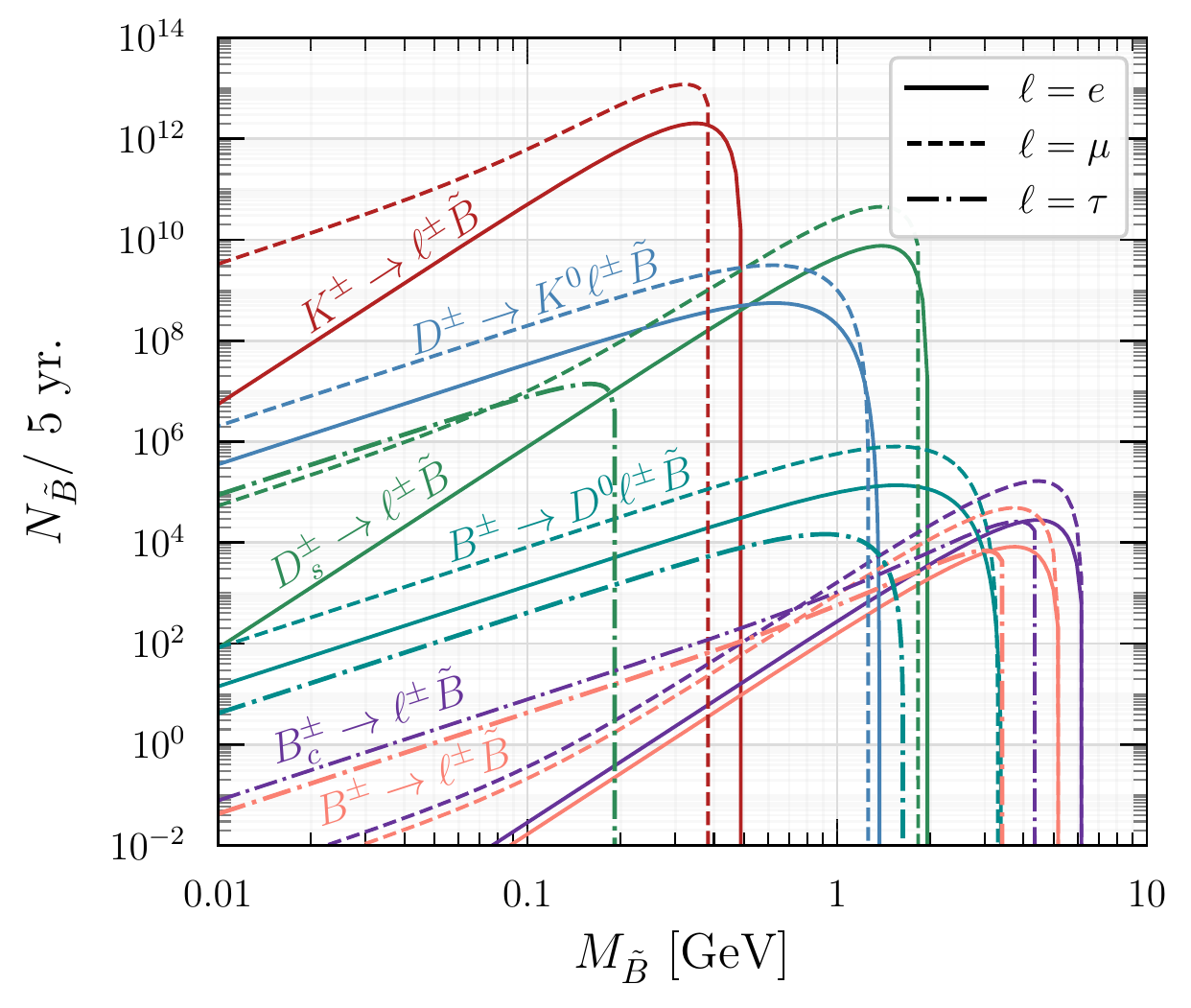}
    \includegraphics[width=0.485\textwidth]{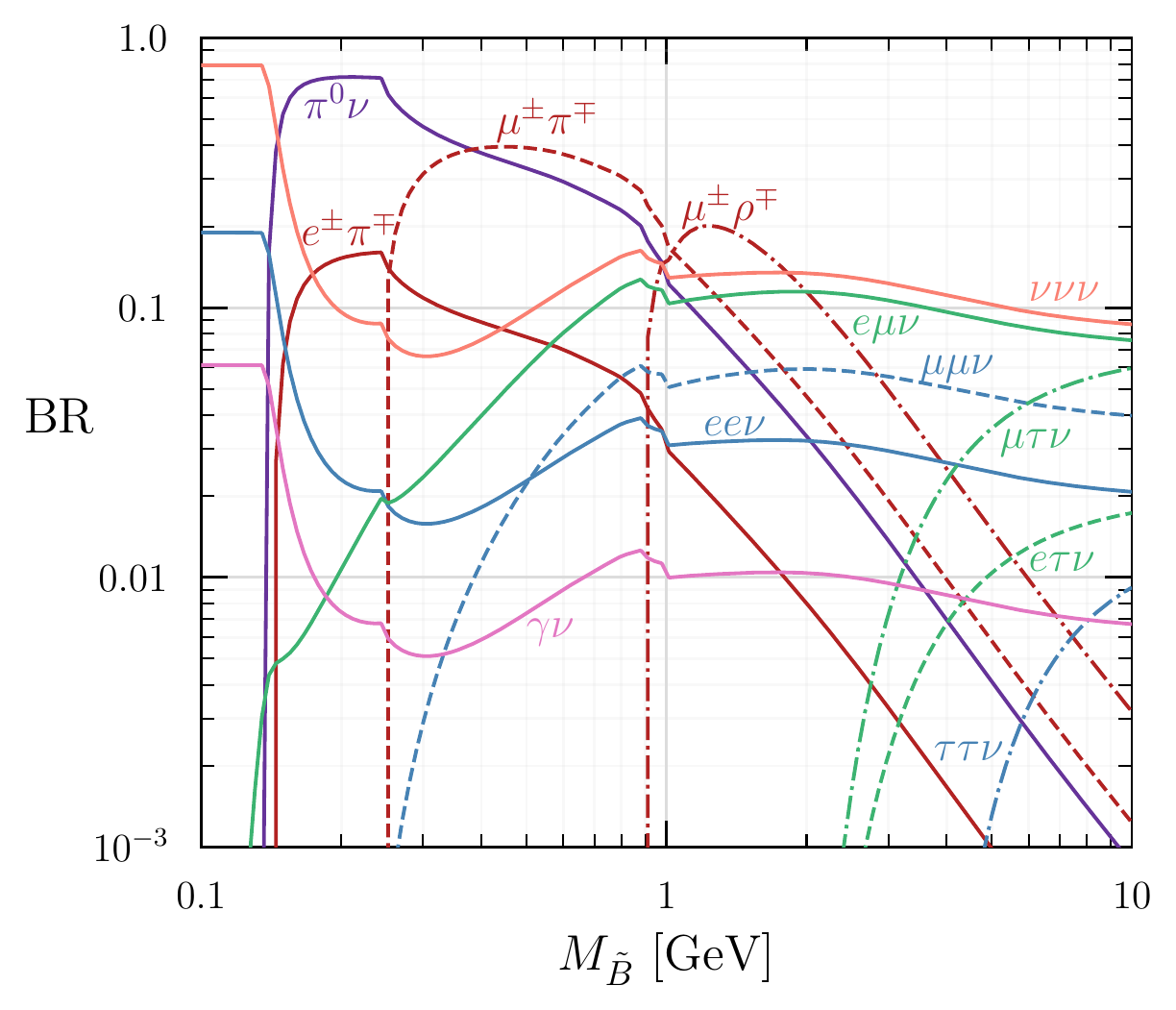}
    \caption{\label{fig:BinoPheno} \textit{Left}: Number of {\bivo}s produced at SHiP assuming 5 years of operation and $\Lambda_M = 1$ TeV. (This messenger scale is taken as a benchmark for comparison to other relevant models.) \textit{Right}: Branching ratios of the {\bivo} into different final states as a function of the {\bivo} mass $M_{\tilde{B}}$.}
\end{figure*}
\section{Bi$\nu$o Phenomenology at SHiP}\label{sec:pheno}
The {\bivo} mixes with the SM neutrinos via interactions given in \eqref{eq:LagMix}. After EWSB, this mixing will induce couplings of the {\bivo} to the EW gauge bosons $W^\pm,Z$, given by
\begin{equation}
\mathcal{L} \supset \frac{g_2}{\sqrt{2}} f_i \frac{M_{\tilde{B}}}{\Lambda_M} W^+_\mu \ell_i \bar{\sigma}^\mu \tilde{B}  + \frac{g_2}{2 \cos\theta_W} f_i \frac{M_{\tilde{B}}}{\Lambda_M} Z_\mu \nu_i \bar{\sigma}^\mu \tilde{B} \,,
\end{equation}
where $g_2$ is the $SU(2)_L$ coupling constant and $\theta_W$ is the Weinberg angle. These interactions will allow the {\bivo} to be produced in any process where a SM neutrino is produced, and to decay directly to EW gauge bosons if $M_{\tilde{B}} > M_{W,Z}$ or to SM fermions via off-shell gauge bosons when $M_{\tilde{B}} < M_{W,Z}$. Since the mixing parameters $f_i$ are fixed by the observed neutrino mass differences, unlike minimal heavy neutral lepton (HNL) scenarios, the {\bivo} phenomenology is completely determined once the {\bivo} mass $M_{\tilde{B}}$ and the messenger scale $\Lambda_M$ are fixed.

\subsection{Bi$\nu$o  Production}

In proton beam dump experiments, {\bivo}s will be copiously produced in two- or three-body decays of mesons in addition to secondary production from the decays of $\tau^\pm$ produced in the decay of $D^\pm_s$ meson. The number of {\bivo}s produced is given by
\begin{equation}\label{NHNL}
N_{\tilde{B}} = N_i \text{BR}(i \to \tilde{B} + X)\,,
\end{equation}
where $N_i$ is the number of mesons,  $N_M$, or $\tau$ leptons, $N_\tau$, produced and BR($i \to \tilde{B} + X$) is the branching ratio for the meson or $\tau^\pm$ decay to a {\bivo} and final states $X$. The full expressions for the branching ratios of meson and $\tau$ decays can be found in \cite{Bondarenko:2018ptm,Coloma:2020lgy}. 

In the left plot of Fig.~\ref{fig:BinoPheno} we show the number of {\bivo}s produced at SHiP assuming 5 years of operation and a messenger scale $\Lambda_M = 1$ TeV. Note that, even though it is excluded by low energy observables, this value of $\Lambda_M$ is chosen for illustration purposes and for ease of translating into other new physics scales. We see that the production rate of the {\bivo} in association with muons (dashed curves) is typically larger than that of electrons (solid curves) and tau leptons (dot-dashed curves) since $f_\mu > f_{e,\tau}$.

\subsection{Bi$\nu$o  Decays}

Once produced, {\bivo}s will decay via the weak interactions into two-body final states, $\tilde{B} \to \ell^\pm M$, $\tilde{B} \to \nu M$, where $M$ is a meson, and three-body final states, $\tilde{B} \to f f^\prime \nu_i$ for final state fermions $f,f^\prime$. The partial widths for the various decays are given in \cite{Gorbunov:2007ak,Atre:2009rg,Bondarenko:2018ptm,Berryman:2017twh,Coloma:2020lgy}. Note that above $M_{\tilde{B}} \sim 1$ GeV, the decay to hadrons is more appropriately described by quark production in the final state. Thus, for $M_{\tilde{B}} <1$ GeV we determine the total hadronic decay rate by summing partial widths for exclusive decays to mesons. Above 1 GeV we switch to the inclusive decay rates to quarks $\tilde{B} \to q q^\prime \nu$, following the approach of \cite{Bondarenko:2018ptm}.

In the right plot of Fig.~\ref{fig:BinoPheno} we show the branching ratios (BRs) as a function of the {\bivo} mass $M_{\tilde{B}}$. The most promising channels to search for are those involving charged particles in the final state. The BRs are independent of the ratio $M_{\tilde{B}}/\Lambda_M$ and depend only on the mixing parameters $f_i$, which are fixed by the neutrino mixing observables. When $M_{\tilde{B}} \lesssim m_\pi$ the most important decay is $\tilde{B} \to ee \nu$, while in the region $m_\pi \lesssim M_{\tilde{B}} \lesssim 1$~GeV the decays $\tilde{B} \to e^\pm \pi^\mp$ and $\tilde{B} \to \mu^\pm \pi^\mp$ become dominant. The decay $\tilde{B} \to \mu^\pm \rho^\mp$ becomes important between 1-2 GeV, while for $M_{\tilde{B}} \gtrsim 2$ GeV the decay $\tilde{B} \to e\mu\nu$ is the most relevant.
\newline

It is important to emphasize that the {\bivo} is directly related to neutrino mass generation, unlike minimal HNL scenarios that can be searched for at SHiP.\footnote{By minimal HNLs we mean the existence of a single HNL that mixes with only one SM neutrino flavor, which is an important benchmark for future experimental searches \cite{Beacham:2019nyx}.} As a result, it is possible to gain insights into the mechanism responsible for neutrino mass generation if a positive signal involving both electrons and muons, for example, are observed. Because the BRs are completely determined by the neutrino mixing parameters, ratios of BRs could tell us if the new particle is involved in neutrino mass generation and would be a smoking gun signal for the model described in Sec.~\ref{sec:model}. We also emphasize that, unlike RPV SUSY models, the relavant {\bivo} phenomenology described here does not depend on sfermion masses. In this way, SHiP will able to probe this model even if the sfermions are much heavier than the LHC capabilities.

\section{SHiP Sensitivity to {\bivo} }\label{sec:sensitiviy}
%
\begin{figure*}[t]
    \centering
        \includegraphics[width=0.585\textwidth]{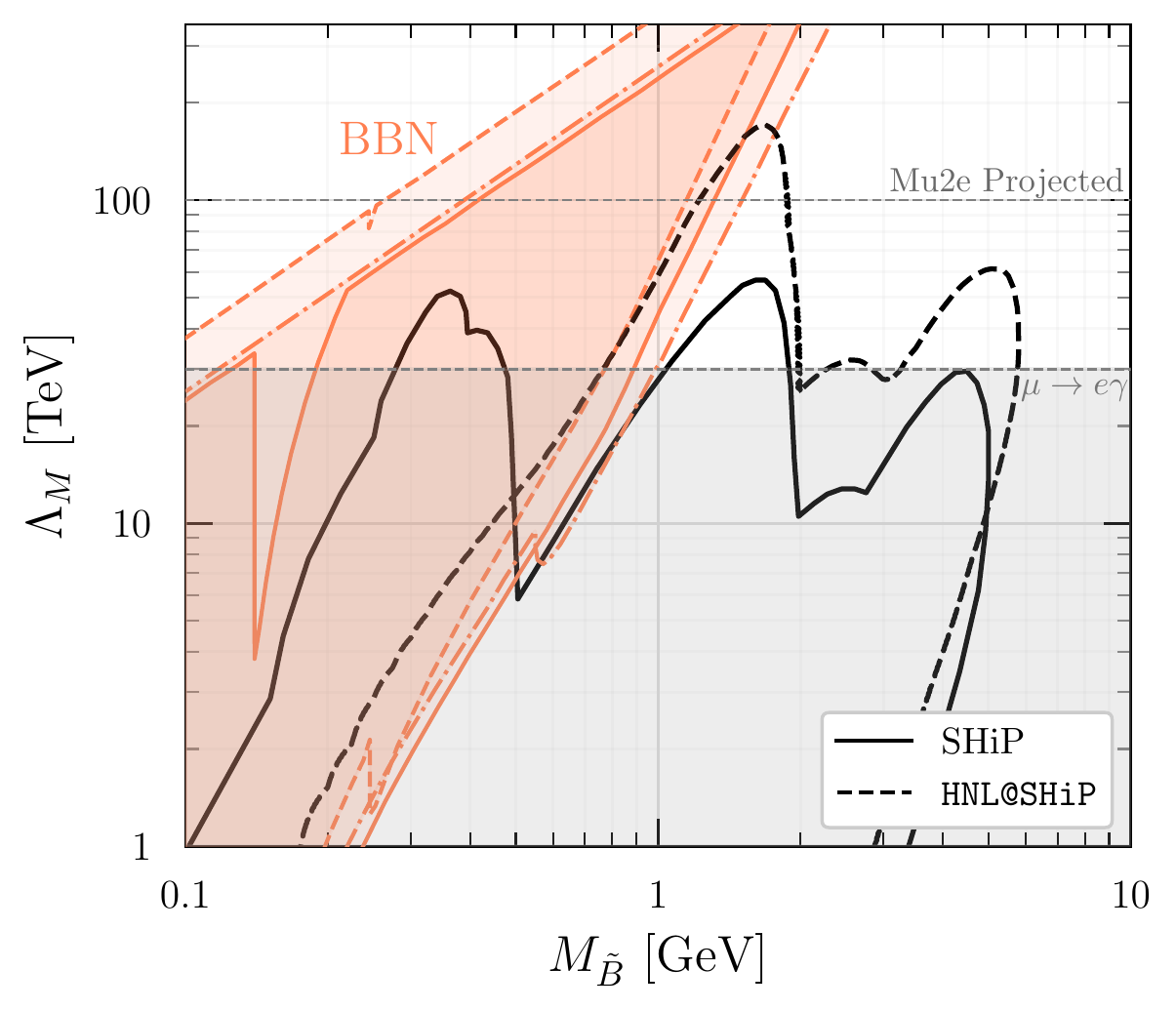}
     \caption{\label{fig:sensitivity} SHiP sensitivity to {\bivo}s assuming 5 years of operation using our conservative method (solid black curve) and the \texttt{HNL@SHiP} method (dashed black curve). The greater sensitivity with the conservative method below $M_{\tilde{B}} \sim 400$ MeV is due to the {\bivo} production from kaons, which is not included in the \texttt{HNL@SHiP} package. Existing limits from $\mu \to e \gamma$~\cite{MEG:2016leq} are depicted by the gray shaded region below the dashed gray line labelled ``$\mu \to e \gamma$". We also show the projected exclusion limit from the proposed Mu2e experiment~\cite{Mu2e:2014fns}, which will look for $\mu \to e$ conversion in nuclei. BBN constraints are shown by the orange shaded regions. See text for further details.}
\end{figure*}
Given the SHiP experimental setup described in Sec.~\ref{sec:Ship}, we can determine the geometrical acceptance and efficiency for detecting the {\bivo} decay products via Monte Carlo simulation.  To determine the geometric acceptance, we require that the visible final states reach the detector after the {\bivo} has decayed anywhere in the decay volume. For a given position $z$ along the decay volume, we require that the final states enter the detector, which has $x$ and $y$ dimensions of 5 m and 10 m, respectively \cite{SHiP:2021nfo}. These conditions can be quantified by the following inequalities \cite{Giffin:2022rei}:
\begin{align}\label{eq:acceptance}
\begin{split}
x_f =& \bigg|\frac{p^{\tilde{B}}_x}{p^{\tilde{B}}_z}z + \frac{p^f_x}{p^f_z}(\ell_\text{sh} + \ell_\text{decay} - z) \bigg| < 2.5 \text{m}\,, \\
y_f =&  \bigg|\frac{p^{\tilde{B}}_y}{p^{\tilde{B}}_z}z + \frac{p^f_y}{p^f_z}(\ell_\text{sh} + \ell_\text{decay} - z)\bigg|  < 5 \text{m}\,,
\end{split}
\end{align}
where  $p^{\tilde{B}}_{x,y,z}$ are the {\bivo} momentum components and $p^{f}_{x,y,z}$ are the final state fermion momentum components. Signal events are chosen to be those  in which the {\bivo} decays into two charged particles in the fiducial decay volume enclosed by $z_\text{min} = \ell_\text{sh} = 64$ m and $z_\text{max} = \ell_\text{sh} + \ell_\text{decay} = 114$ m, and satisfy the conditions in \eqref{eq:acceptance}.

The geometric acceptance depends on the location where the {\bivo} decays. Therefore, the total efficiency is an integral over the length of the decay volume given by \cite{Giffin:2022rei}
\begin{equation}\label{eq:eff}
\text{eff} = M_{\tilde{B}}\Gamma \int_{z_\text{min}}^{z_\text{max}} dz \sum_{\text{events}~ \in~\text{geom.}} \frac{e^{-zM_{\tilde{B}}\Gamma/p_z}}{N_\text{MC} p_z}\,,
\end{equation}
where $M_{\tilde{B}}$, $\Gamma$, and $p_z$ are the mass, decay width, and $z-$component of the momentum of the {\bivo}, respectively. The sum is over the events that fall within the geometric acceptance of the detector, and $N_\text{MC}$ is the total number of simulated events. Efficiency plots for minimal HNLs can be found in \cite{Batell:2020vqn,Giffin:2022rei}, for example.

Given the production, decay, and efficiency information discussed above, the total number of signal events is given by
\begin{equation}\label{eq:Nsignal}
N_\text{sig} = N_{\tilde{B}} \times \text{BR}(\tilde{B} \to i) \times \text{eff}_i\,,
\end{equation}
where $N_{\tilde{B}}$ is the number of {\bivo}s produced in a particular production mode, BR($\tilde{B} \to i$) is the branching ratio for $\tilde{B}$ decaying to a final states $i$, and eff$_i$ is the efficiency for detecting the final states. Here we assume that the detector efficiency for reconstructing the charged final states is 100\%. 

To determine the reach of SHiP to GeV-scale {\bivo}s we use the median expected exclusion significance \cite{Cowan:2010js,Kumar:2015tna}:
\begin{equation}\label{eq:Zexcl}
Z_\text{excl} = \sqrt{2[s-b \ln(1+s/b)]}\,,
\end{equation}
where $s$ and $b$ are the number of signal and background events, respectively. For 5 years of operation $b=0.1$ background events are expected at SHiP \cite{Ahdida:2654870}.  To set 95\% C.L. exclusion limits we require $Z_\text{excl} > 1.645$, which corresponds to $s \approx 2$ signal events.

We use two methods to determine the SHiP sensitivity to GeV-scale long-lived {\bivo}s. \textbf{(i)} In the first method, we perform a Monte Carlo simulation to simulate the production and decay of the {\bivo}. For the production, we use \texttt{Pythia8} \cite{Sjostrand:2006za,Sjostrand:2014zea} to simulate a 400 GeV proton beam striking a proton at rest, and extract the four momenta of mesons that are produced. We use this output to determine the lab frame momentum of the {\bivo}. 

For two-body decays of the {\bivo}, we can analytically solve for the final state momenta of its decay products in the rest frame of the {\bivo}. For three-body decays, we use the the publicly available code \texttt{muBHNL} which uses the differential decay distributions of HNLs to generate a weighted sample of final state momenta in the rest-frame of the HNL \cite{Kelly:muBHNL,deGouvea:2021ual,Kelly:2021xbv}. Note that, for this purpose, a {\bivo} with GeV-scale mass is a type of HNL with generic mixing with SM neutrinos, and the kinematic output of \texttt{muBHNL} is directly applicable.

For both two- and three-body decays, we then use the lab-frame momentum of the {\bivo} to boost the final state momenta into the lab frame. With the lab-frame momenta of the {\bivo} and its decay products, we can determine the geometric acceptance and efficiency using \eqref{eq:eff}, calculate the signal rate using \eqref{eq:Nsignal}, and find the 95\% C.L. exclusion limits from \eqref{eq:Zexcl}. 

For each {\bivo} mass, we consider only the production and decay mode which leads to the best sensitivity, rather than a combination of all possible production and decay modes. These exclusion limits can thus be considered to be relatively conservative. 

\textbf{(ii)} The second method we employ to obtain exclusion limits is by using the \texttt{Mathematica} based package \texttt{HNL@SHiP} \cite{HNLatSHIP} that determines the SHiP sensitivity to HNLs with arbitrary mixings with SM neutrinos; see \cite{SHiP:2018xqw} for more details. The \texttt{HNL@SHiP} package combines all production and decay modes for a given {\bivo} mass, and the exclusion limits with this method can be considered to be much more aggressive compared to the first method described above. 

Another important difference between the two methods is the omission of kaons in the \texttt{HNL@SHiP} package. Initially, it was expected that the interaction length of kaons in the SHiP beam dump is much shorter than their decay length and that the kaons would be absorbed before decaying. However, it was shown in \cite{Gorbunov:2020rjx} that the kaon production rate can be substantial. We've included {\bivo} production from kaons in method \textbf{(i)}, which sets the leading SHiP constraints below $M_{\tilde{B}} \simeq 400$ MeV.

\section{Results and Discussion}

The exclusion limits derived using these two analyses are depicted in Fig.~\ref{fig:sensitivity}. The solid curve corresponds to the more conservative analysis, where we only consider the channels that lead to the best sensitivity rather than a combination of all production and decay channels. The dashed black curve depicts the exclusion limits derived using the \texttt{HNL@SHiP} code. 

Additionally, we show current and future constraints from $\mu \to e \gamma$~\cite{MEG:2016leq} and $\mu \to e$ conversion in nuclei~\cite{Mu2e:2014fns} by the horizontal dashed gray lines labeled ``$\mu \to e \gamma$'' and ``Mu2e Projected'', respectively.  The current bound of BR($\mu \to e\gamma) < 4.2\times10^{-13}$ results in a lower bound on the messenger scale $\Lambda_M \gtrsim 35$ TeV, while future projections of the sensitivity of the Mu2e experiment to $\mu \to e$ conversion near nuclei is expected to probe upto $\Lambda_M \lesssim 100$ TeV. These constraints are independent of the {\bivo} mass.

Finally, the shaded orange regions are constraints from BBN. To obtain these bounds, we assume that the {\bivo} mixes with a single neutrino flavor  $\nu_i$ with a mixing parameter $U_i \sim f_i M_{\tilde{B}}/ \Lambda_M$, with $f_i$ given by \eqref{eq:mixing}. We then use the results of \cite{Sabti:2020yrt,Boyarsky:2020dzc} for BBN constraints on HNLs that mix with a single neutrino flavor to constrain the {\bivo} parameter space. This is depicted by the three orange shaded regions. The regions enclosed by the solid, dashed, and dot-dashed orange curves are BBN constraints for electron-, muon-, and tau-mixed HNLs, respectively. For the {\bivo}, which mixes with all three neutrino flavors, BBN constraints will  generally lie in between the dashed and the dot-dashed curves enclosing the upper-left and lower-right orange shaded region. 

We observe that already with our conservative analysis (the solid black curved in Fig.~\ref{fig:sensitivity}) SHiP can begin to probe the $M_{\tilde{B}} = 1-2$ GeV window that is not excluded by the current $\mu \to e \gamma$ and BBN limits, with messenger scales up to $\Lambda_M \sim 60$ TeV. Combining all possible production and decay modes is expected to lead to better exclusion limits allowing SHiP to probe the mass region above the $D^\pm_s$ meson mass for the messenger scales not excluded by current limits. We also see that SHiP can even probe messenger scales beyond the projected sensitivity of the Mu2e experiment. 

These results show that SHiP is highly complementary to experiments looking for charged lepton flavor violation, as well as constraints from BBN for the {\bivo} model. SHiP is able to probe new parameter space for {\bivo} masses in the $M_{\tilde{B}}\sim1-5$ GeV range, corresponding to a messenger scale in the $\Lambda_M \sim 60-200$ TeV range.

\section{Conclusion}\label{sec:conclusion}
In this paper, we investigated the sensitivity of the SHiP experiment to the bino mass and the SUSY messenger scale in a $U(1)_{R-L}$ extension of the MSSM which explains the origin of neutrino masses. (In this scenario, the bino and its Dirac partner singlino act like right-handed neutrinos and are responsible for generating the neutrino masses.) Previous studies have considered the production of {\bivo}s at the LHC from the decays of squarks, and constrained  a combination of the squark mass, the {\bivo} mass, and the messenger scale $\lambda_M$. However, if the squarks are too heavy to be produced at the LHC, weak interactions might be too weak to produce {\bivo}s at the LHC.

On the other hand, if the {\bivo} mass is in the MeV-GeV range, a large production rate can be achieved in high energy beam dump experiments. Mixing with SM neutrinos allows the {\bivo} to be produced from meson decays, which have large production rates at SHiP. We showed that SHiP can probe messenger scales up to 200~TeV and probe a parameter region complementary to experiments looking for charged-lepton flavor violation.

The model we investigated is directly related to neutrino mass generation that can be discovered at SHiP. If final states involving both muons and electrons are observed, their relative widths are fully predicted by neutrino-mass-mixing measurements. This would provide an incredible opportunity to search for the explanation of one of the most pressing problems with the SM.

\section*{Acknowledgements}
This work is supported in part by the Natural Sciences and Engineering Research Council (NSERC) of Canada. The work of DT is supported by the Arthur B. McDonald Canadian Astroparticle Physics Research Institute.

\appendix


\bibliography{references}

\end{document}